\documentclass[10pt,final, twocolumn]{IEEEtran}
\IEEEoverridecommandlockouts
\def\BibTeX{{\rm B\kern-.05em{\sc i\kern-.025em b}\kern-.08em
    T\kern-.1667em\lower.7ex\hbox{E}\kern-.125emX}}

\usepackage{amsmath,algorithm, algorithmic}
\usepackage{cite}
\usepackage{epsf,graphics,graphicx}
\usepackage{comment}
\usepackage{bm}
\usepackage{amssymb,amsthm}
\usepackage{amsmath,flushend}
\usepackage[top=.75in, left= 0.625in, right = 0.625in, bottom =1in]{geometry}

\newtheorem{Proposition}{Proposition}
\newtheorem{remark}{Remark}
\newtheorem*{remark*}{Remark}

\usepackage{comment}
\usepackage[textwidth=30mm]{todonotes}
\usepackage{soul}

\setstcolor{magenta}
\sethlcolor{lightmauve}

\definecolor{deepmagenta}{rgb}{0.8, 0.0, 0.8}
\definecolor{lightmauve}{rgb}{0.86, 0.82, 1.0}
\definecolor{green-yellow}{rgb}{0.68, 1.0, 0.18}
\definecolor{lightskyblue}{rgb}{0.53, 0.81, 0.98}
\definecolor{beaublue}{rgb}{0.74, 0.83, 0.9}

\newcommand{\Revadd}[1]{\textcolor{black}{#1}}

\def\vec#1{{\bf #1}}
\def\Tr#1{\mathrm{Tr}\left\{ #1 \right\} }
\def\outer{\otimes}

\DeclareMathOperator{\ex}{\mathbb{E}}

\newcommand{\ZZ}{\mathbb{Z}}

\def\boldmu{\mbox{\boldmath$\mu$}}
\def\boldtau{\mbox{\boldmath$\tau$}}

\def\bLambda{\mbox{\boldmath$\Lambda$}}
\def\bPhi{\mbox{\boldmath$\Phi$}}
\def\bSigma{\mbox{\boldmath$\Sigma$}}

\def\bG{{\bf G}}

\def\bI{{\bf I}}
\def\bQ{{\bf Q}}
\def\bR{{\bf R}}
\def\bS{{\bf S}}
\def\bT{{\bf T}}

\def\bV{{\bf V}}
\def\bW{{\bf W}}

\def\bY{{\bf Y}}
\def\bZ{{\bf Z}}

\def\bZero{{\bf 0}}

\def\nt{{N_t}}
\def\nr{{N_r}}
\def\ns{{N_s}}

\def\nue{{M}} 
\def\nris{K} 

\def\snrdir{{\gamma_{dm}}} 

\def\bRcal{\mathbfcal{R}}
\def\bTcal{\mathbfcal{T}}

\DeclareMathAlphabet\mathbfcal{OMS}{cmsy}{b}{n}

\def\bk{{\bf k}}

\def\bq{{\bf q}}

\def\bs{{\bf s}}

\def\bu{{\bf u}}
\def\bv{{\bf v}}

\def\bx{{\bf x}}
\def\by{{\bf y}}
\def\bz{{\bf z}}

\begin{document}
\bstctlcite{BSTcontrol}

\title{MIMO MAC Empowered by\\ Reconfigurable Intelligent Surfaces:\\ Capacity Region and Large System Analysis}

\author{Aris~L.~Moustakas,~\IEEEmembership{Senior~Member,~IEEE,} and George~C.~Alexandropoulos,~\IEEEmembership{Senior~Member,~IEEE}
\thanks{This work has been supported by the Smart Networks and Services Joint Undertaking (SNS JU) project TERRAMETA under the European Union’s Horizon Europe research and innovation programme under Grant Agreement No 101097101, including top-up funding by UK Research and Innovation
(UKRI) under the UK government’s Horizon Europe funding guarantee.}
\thanks{A.~L.~Moustakas is with the Department of Physics, National and Kapodistrian University of Athens, 15784 Athens, Greece and with the Archimedes/Athena Research Unit, Athens, Greece (e-mail: arislm@phys.uoa.gr).}
\thanks{G.~C.~Alexandropoulos is with the Department of Informatics and Telecommunications, National and Kapodistrian University of Athens, 15784 Athens, Greece and with the Department of Electrical and Computer Engineering, University of Illinois Chicago, Chicago, IL 60601, USA (e-mail: alexandg@di.uoa.gr).}
}

\maketitle

\begin{abstract}
Smart wireless environments enabled by multiple distributed Reconfigurable Intelligent Surfaces (RISs) have recently attracted significant research interest as a wireless connectivity paradigm
for sixth Generation (6G) networks. In this paper, using random matrix theory methods, we calculate the mean of the sum Mutual Information (MI) for the correlated Multiple-Input Multiple-Output (MIMO) Multiple Access Channel (MAC) in the presence of multiple RISs, in the large-antenna number   limit. 
We thus    obtain the capacity region boundaries, after optimizing over the tunable RISs' phase configurations.  
Furthermore, we obtain a closed-form expression for the variance of the sum-MI metric, which together with the mean provides a tight Gaussian approximation for the outage probability. The derived results become relevant in the presence of fast-fading, when channel estimation is extremely challenging. Our numerical investigations showcased that, when the angle-spread in the neighborhood of each RIS is small, 
which is expected for higher carrier frequencies, the communication link strongly improves from optimizing the ergodic MI of the multiple RISs. We also found that, increasing the number of transmitting users in such MIMO-MAC-RIS systems results to rapidly diminishing sum-MI gains, hence, providing limits on the number of users that can be efficiently served by a given RIS.
\end{abstract}


\section{Introduction}
The upcoming sixth Generation (6G) of wireless networks is envisioned to connect the physical, digital, and human worlds, enabling ubiquitous wireless intelligence and extremely demanding communications, sensing, and computing applications (e.g., holographic imaging~\cite{Gong_HMIMO_2023}, digital twinning~\cite{Masaracchia_DT_2023}, and edge artificial intelligence~\cite{Letaief_edgeAI_2023}). This vision necessitates advances at various aspects of the network design and at the orchestration of its diverse components, as well as novel multi-sensory device technologies. Smart wireless environments, enabled by the technology of Reconfigurable Intelligent Surfaces (RISs)~\cite{di2019smart,RIS_challenges,WavePropTCCN,risTUTORIAL2020,pan2022overview,zhi2022active,single_amplifier_george,RIS_Scattering}, constitute a recent revolutionary wireless connectivity paradigm contributing to the latter vision, according to which, the propagation of information-bearing electromagnetic signals can be dynamically programmed over the air in a cost-, power-, and computationally-efficient manner~\cite{RISE6G_COMMAG}. 

Recent theoretical investigations on the joint optimization of massive Multiple-Input Multiple-Output (MIMO) transceivers with RISs have demonstrated the potential of RIS-enabled smart wireless environments for improving spectral and energy efficiencies~\cite{huang2019reconfigurable,KSI+22,KDA22}, localization accuracy~\cite{RIS_Localization}, computing~\cite{RIS_Computing_2023}, physical-layer security~\cite{PLS_Kostas}, as well as the integration of sensing and communications~\cite{RIS_ISAC_SPM}. More specifically, in~\cite{YXH+21}, focusing on a single-RIS-assisted multi-user MIMO system in the uplink direction and considering slowly varying statistical Channel State Information (CSI) availability between the user terminals and the RIS, an algorithm for the joint passive and active beamforming optimization was proposed. A multi-user MIMO system aided by multiple RISs, with the objective to maximize the system sum rate with respect to the RISs' phase configurations and the beamforming vectors at the Transmitters (TXs), was studied in \cite{ADD21}. 
In \cite{Yang_2021c}, the problem of energy efficiency optimization for a wireless communication system assisted by multiple distributed RISs was investigated. RIS-empowered Device-to-Device (D2D) communications underlaying a cellular network were considered in~\cite{Yang_2021d}, in which an RIS was missioned to enhance the desired signals and suppress interference between paired D2D and cellular links. Targeting the uplink achievable sum-rate maximization in a D2D cellular system with multiple distributed RISs deployed at the cell boundaries in~\cite{CLN+21}, a low-complexity decentralized optimization algorithm was presented. In~\cite{YCS+22}, a system with multiple distributed RISs was investigated for energy efficiency, optimizing the dynamic control of the on-off status of each RIS as well as the phase profiles of the on-state RISs. An RIS-aided multi-user MIMO system with statistical CSI availability, aiming at sum-rate maximization, was studied in~\cite{ZMS+23}, while~\cite{CHP+23} focused on optimizing the uplink of an RIS-assisted wideband multi-user system operating in the near-field regime. Multi-RIS-enabled intelligent wireless channels were also studied in \cite{Samarakoon_2020,pervasive_DRL_RIS}, with their optimization being performed using, respectively, supervised and unsupervised artificial intelligence methodologies.

Another category of research works on RIS-enabled smart wireless environments deals with the characterization of their fundamental capacity limits. In particular,~\cite{Mu_2021b,LSC+23} focused on the downlink of an RIS-assisted multi-user wireless communication system, with the latter work considering millimeter-wave channels under the assumption of statistical CSI availability. In~\cite{PTR+22}, the authors analyzed the achievable sum rate of the RIS-aided MIMO broadcast channel showcasing the advantages of deploying multiple RISs, and presented optimization algorithms for the passive and active beamforming optimization. \Revadd{More recently, the authors in \cite{chen2023fundamental} took advantage of the duality between uplink and downlink to characterize the dirty-paper-coding capacity region of a multi-antenna broadcast TX to multiple single-antenna Receivers (RXs) in the presence of a single RIS, and optimized the system via a less complex zero-forcing transmission scheme.} An asymptotic analysis of the uplink data rate in a multi-user setup with a single RIS acting as a receiver was carried out in~\cite{Jung2020}, considering spatially correlated Rician fading channels subject to estimation errors as well as RIS hardware impairments. For the same system model, but for the downlink direction,~\cite{Nadeem2020} studied the optimum linear precoding matrix that maximizes the minimum signal-to-interference-plus-noise ratio, and presented deterministic approximations for its parameters. An asymptotic analysis and approximations for the interference-to-noise ratio considering uncorrelated Rayleigh fading channels were derived in~\cite{INR_analysis}, while~\cite{analysis_intertwinement} presented upper and lower bounds for the outage probability and ergodic capacity, considering the intertwinement model of 
\cite{Abeywickrama_2020_all} for the amplitude and phase relation in the RIS element response. In 
\cite{Gao2021_irs_train_outage}, the authors generalized the performance metric to the outage probability, in the context of high speed trains. The achievable sum capacity of an RIS-aided multi-user MIMO system was also analyzed in~\cite{JS23} and two efficient transmission schemes were designed. The effect of a single RIS on the uplink transmission from multiple TXs to a single multi-antenna RX was analyzed in~\cite{You_2021}, obtaining an asymptotic expression for the sum Mutual Information (MI) averaged over the channel matrices only close to the RIS, and thus, neglecting the fast-fading effects close to the TXs and RX. To the best of our knowledge,~\cite{Moustakas2023_RIS} and its conference version were the first works analyzing the asymptotic outage capacity of a multi-RIS-enabled smart wireless environment assisting a point-to-point Kronecker-correlated MIMO system.

\subsection{Contributions}
In this paper, we study the capacity of MIMO-MAC-RIS communication systems, comprising multiple multi-antenna TXs simultaneously communicating with a single multi-antenna RX in the presence of multiple RISs. We present novel closed-form expressions for the asymptotic statistics of the sum-MI performance, which are used to assess the potential gains from the RISs, but also their limitations. A summary of the contributions of this work appear below:
\begin{itemize}
\item
Applying methods used originally in the context of statistical physics, we obtain a closed-form expression for the average sum-MI with arbitrary relative priorities for each multi-antenna TX in the presence of multiple RISs. This expression, when optimized over the phase configuration matrices of the RISs, provides the capacity region for the MIMO-MAC-RIS problem. This expression is nominally valid in the limit of large numbers for the TXs/RX antenna elements and large numbers for the phase-tunable elements of the RISs.

\item Using the above methods, we derive an analytic expression for the variance of the sum-MI. We discuss why the higher cumulant moments of the distribution of the sum-MI vanish in the asymptotic limit, as a result of which the distribution of the sum-MI converges weakly to the Gaussian distribution. Thus we obtain an approximation for its outage probability for block-fading channels. This approximation is corroborated using Monte Carlo simulations, where the agreement with the approximate Gaussian distribution is exceptional down to outage probabilities of $10^{-3}$ even for relatively small antenna arrays at both the TXs and RX.
\item By considering the ergodic sum-MI as a metric of the MIMO-MAC-RIS system performance, the phase configurations of the multiple RISs are designed via two novel approaches, a semi-optimal and an optimal one, assuming that only the statistical properties of the involved channels are known, which is more realistic, given their slow variations compared to the fading of the channels. The optimization with respect to the RISs' reflections using the analytic expressions of the ergodic sum-MI performance is far more efficient compared to optimizing the instantaneous sum-MI. Furthermore, due to a decoupling effect, the optimization over the multiple RISs can be performed at separate steps, hence, it can be easily parallelized.
\item Using the derived ergodic sum-MI expression, we analyze the optimization gains over the phase configurations of the multiple RISs and compare the throughput of the optimal solution to the semi-optimal one. Furthermore, we assess the effectiveness of RISs' phase optimization as the number of TXs increases and observe that beyond a certain point the relative gains vanish. 
\item We compare our numerically evaluated analytic results with Monte Carlo simulations, demonstrating very good agreement even though we use  relatively small antenna arrays at the TXs and RX. \Revadd{We also evaluate the effect of phase quantization of the RIS tunable elements to the overall performance of the MIMO-MAC-RIS system. It is showcased that, even with an $1$-bit phase quantization, the optimization gains are significant, while with a $2$-bit quantization, the performance is nearly optimal.}
\end{itemize}

\subsection{Outline}
Section~\ref{sec:MIMO channel model} describes the proposed system and channel models, while Section~\ref{sec:MI_Analysis} introduces the closed-form expressions for the first two cumulant moments of the sum-MI and the  MIMO-MAC-RIS capacity region, delegating a sketch of the proof for Appendix~\ref{app:proof_ergMI}.
Section~\ref{sec:MI_Optimization} showcases the optimization methodology of the asymptotic sum-MI performance and capacity regions as a function of the phase matrices of the multiple RISs. In Section~\ref{sec:Numerical_Results}, we introduce a number of different situations, comparing our analytic results with numerically generated ones after optimizing over the phase matrices. Finally, in Section~\ref{sec:conclusion}, we conclude and discuss future directions. 

\subsection{Notations} 
Bold-faced upper-case letters will denote matrices, e.g., $\vec X$, with its $(i,j)$-element expressed as $[\vec X]_{i,j}$, while bold-faced lower-case letters will stand for column vectors, e.g., $\vec x$, with its $i$-element given by $[\vec x]_i$. The superscripts $T$ and $\dagger$ denote
transposes and Hermitian conjugates, $\Tr{\, \cdot\,  }$ and $\left\Vert\cdot\right\Vert_F$ stand for the trace and Frobenius norm of a matrix, respectively, and $\vec I_n$ ($n\geq2$) expresses to the $n$-dimensional identity matrix. The superscripts/subscripts $t$ and $r$ denote quantities (e.g., channel matrices) corresponding to the TX and RX, respectively. $\mathbf{x}\sim{\cal CN}(\mathbf{0}_n,\mathbf{I}_n)$ expresses a $n$-element complex and circularly symmetric Gaussian vector with zero-mean elements and covariance matrix $\mathbf{I}_n$, while $\ex[\,\cdot\,]$ is the expectation operator.

\section{System and Channel models}\label{sec:MIMO channel model}
\begin{figure}[!t]
	\centering
	\includegraphics[width=\columnwidth]{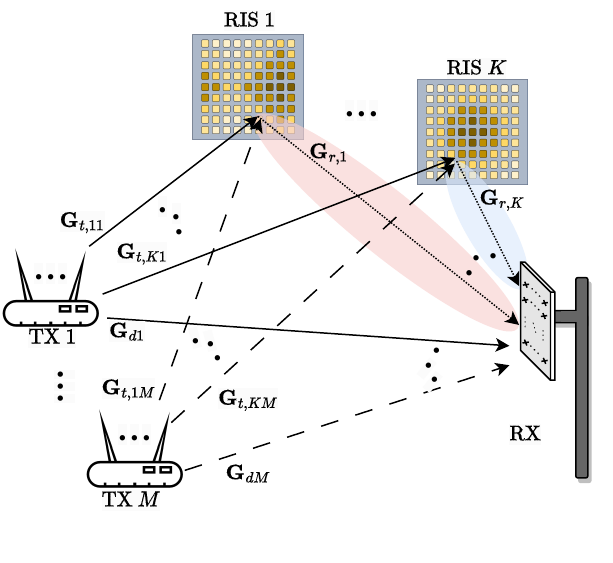} 
	  \caption{
   The considered MIMO-MAC-RIS communication system comprising $\nue$ TXs each equipped with $\nt$ antenna elements, a single $\nr$-antenna RX, and $K$ identical RISs. Notations $\bG_{\alpha,\beta}$, detailed in Section~\ref{sec:MIMO channel model}, represent the channel gain matrices between any pair $\alpha$ and $\beta$ of the latter network nodes.
   }
		\label{fig:system_model}
\end{figure}

The considered wireless uplink communication system comprises $\nue$ Transmitters (TXs), each equipped with $\nt$ antenna elements, simultaneously communicating with a single $\nr$-antenna RX via the assistance of $\nris$ identical RISs, each consisting of $\ns$ tunable reflecting elements~\cite{Tsinghua_RIS_Tutorial}, as depicted in Fig.~\ref{fig:system_model}. The fading channel under which this MIMO-MAC-RIS system operates includes the $\nue$ direct TX-RX links as well as channel components resulting from the programmable reflections due to the multiple RISs. All channel gains are assumed to be perfectly available at the RX via an adequate estimation approach (e.g.,  \cite{Tsinghua_RIS_Tutorial,hardware2020icassp,Swindlehurst_CE,HRIS_CE_all,zhi2022power}), but not to any of the TXs. The $\nr$-dimensional received baseband signal vector at the RX is given by the following expression:
\begin{equation}\label{eq:basic_channel_eq}
  \by \triangleq \sum_{m=1}^\nue \bG_{{\rm tot},m}  \bx_m + \bz,
\end{equation}
where $\bz\sim{\cal CN}(\mathbf{0}_\nr,\mathbf{I}_\nr)$ represents the thermal noise vector and $\bx_m$ with $m=1,2,\ldots,M$ is the $\nt$-dimensional signal vector transmitted from the $m$-th TX having signal covariance matrix $\bQ_m\triangleq \ex[\bx_m\bx_m^\dagger]$, which has the following normalization $\Tr{\bQ_m}=\rho_m\nt$.  $\rho_m$ denotes the Signal-to-Noise Ratio (SNR) of the $m$th TX, assumed henceforth for simplicity to be the same for all TXs, i.e., $\rho_m=\rho$ $\forall m$.
Each $\nr\times\nt$ matrix of channel amplitudes $\bG_{{\rm tot},m}$  in \eqref{eq:basic_channel_eq} between RX and each $m$-th TX can be expressed as follows:
\begin{align}\label{eq:Gtot}
\bG_{{\rm tot},m} \triangleq \bG_{dm} + \sum_{k=1}^\nris\bG_{r,k}\bPhi_k\bG_{t,km},
\end{align}
with $\bG_{r,k}$ and $\bG_{t,km}$, for $k=1,2,\ldots,\nris$, denoting the $\nr\times \ns$ and $\ns\times\nt$ channel matrices between the RX and each $k$-th RIS and between that $k$-th RIS and each $m$-th TX, respectively. $\bG_{dm}$ is  the $\nr\times\nt$ matrix with elements the direct channel amplitudes between the RX and each $m$th TX; the latter channel does not include any reflection from any of the RISs. In addition, each $\bPhi_k$ denotes the $\ns$-dimensional diagonal square matrix containing the tunable reflection coefficients at each $k$-th RIS in the main diagonal. 

The diagonal elements of the phase matrices $\bPhi_k$'s are modeled to have unit norm \cite{risTUTORIAL2020}, with each $n$-th reflection coefficient, with $n=1,2,\ldots,\ns$, of each $k$-th RIS expressed as $[\bPhi_k]_{n,n}\triangleq e^{i\phi_{k,n}}$. 
\Revadd{In this paper, we have modeled the characteristics of the channel matrices using the so-called Kronecker-product correlation model. While this model shares similarities with the Rician fading channel model, in that it can model line-of-sight channels using small angular spreads, it is more flexible in the sense that it can provide different correlation levels at the communication ends of the channel.} Thus, all channel matrices in $\bG_{{\rm tot},m}$ are assumed complex Gaussian, independent for different values of $m$, and having the Kronecker-product-type covariances  $\forall i,j=1,2,\ldots,\nr$, $\forall \ell,n=1,2,\ldots,\nt$, and $\forall a,b=1,2,\ldots,\ns$:
\begin{subequations}\label{eq:all_correlations}
\begin{align}
\label{eq:Gd_cov}
\ex\left[[\bG_{dm}]_{i,\ell}[\bG_{dm}]_{j,n}^*\right]&=\frac{\snrdir}{\nt}[\bR_{dm}]_{i,j} [\bT_{dm}]_{\ell,n},
\\ \label{eq:Grk_cov}
\ex\left[[\bG_{r,k}]_{i,a}[\bG_{r,k}]_{j,b}^*\right]&=\frac{1}{\nt}[\bR_{k}]_{i,j} [\bS_{r,{k}}]_{a,b},
\\ \label{eq:Gtk_cov}
\ex\left[[\bG_{t,km}]_{a,\ell}[\bG_{t,km}]_{b,n}^*\right]&=\frac{1}{\nt}[\bS_{t,km}]_{a,b} [\bT_{km}]_{\ell,n}.
\end{align}
\end{subequations}
The signal-impinging-side correlation matrices in the latter expressions, namely $\bR_{k}$, $\bS_{r,k}$, and $\bR_{dm}$, as well as the outgoing-signal-side correlation matrices, namely $\bT_{km}$, $\bS_{t,km}$, and $\bT_{dm}$ are all non-negative definite, with traces  fixed to: $\Tr{\bT_{km}}=\Tr{\bT_{dm}}=\nt$, $\Tr{\bR_{k}}=\Tr{\bR_{dm}}=\nr$, and $\Tr{\bS_{r,k}}=\Tr{\bS_{t,km}}=\ns$. Note that elements of the matrix $\bS_{t,km}$ represent the correlation coefficient between the incoming EM waves from the $m$th TX at the 
elements of the $k$-th RIS,
while the elements of $\bS_{r,k}$ denote the correlation between outgoing (reflected) EM waves from the $k$th RIS to the RX. Finally, $\snrdir$ represents the ratio of the SNR of the direct link between each $m$-th TX and RX over the SNR of the link between the same nodes realized via the $k$-th RIS.
In this paper, EM waves are treated for simplicity as scalars; the extension to polarized EM-waves will be treated in a future work.

Similar to \cite{Moustakas2000_BLAST1_new,Moustakas2023_RIS}, each $(a,b)$-element of each correlation matrix $\bS_{t,km}$ in \eqref{eq:all_correlations} can be obtained as follows:
\begin{equation}\label{eq:corr_mat_w(k)_def}
    \left[\bS_{t,km}\right]_{ab}=\int \, w_{t,km}(\bk) e^{i\bk^T(\bx_a-\bx_b)}d\Omega_{\bk},
\end{equation}
where $w_{t,km}(\bk)$ represents a normalized weight function of power of the incoming EM wave from direction $\bk$, which denotes the corresponding $3$-dimensional wave vector with magnitude $|\bk|=k_0\triangleq\frac{2\pi}{\lambda}$, where $\lambda$ is the wavelength. Note that this function can be characterized by the mean direction of arrival $\bs_0$ (with $|\bs_0|=k_0$) and the Angle Spread (AS) $\sigma$ (in radians) via the following expression~\cite{Moustakas2023_RIS}:
\begin{equation*}\label{eq:weight_fn_def}
    w_{t,km}(\bk)\propto e^{-\frac{|\bk-\bs_0|^2}{2\sigma^2k_0^2}}.
\end{equation*}
The three-dimensional vectors $\bx_a$ and $\bx_b$ in \eqref{eq:corr_mat_w(k)_def} denote the positions of the meta-material elements $a$ and $b$ within the $k$-th RIS. The integral in this expression is evaluated over the whole unit sphere, and it is normalized so that for $\bx_a=\bx_b$ it gives unity. A similar expression can be given for all correlation matrices in \eqref{eq:all_correlations}.

\section{MIMO-MAC-RIS Capacity Analysis}
\label{sec:MI_Analysis}
Our aim in this paper is to characterize wireless communications between multiple TXs and a single RX in the presence of RISs. It is therefore important to analyze the effect that RISs have for MIMO-MAC systems. In the following analysis, we assume as usual that the RX knows the overall end-to-end channel matrices $\bG_{{\rm tot},m}$ $\forall$$m$ given in \eqref{eq:Gtot} (through, e.g., pilot-assisted channel estimation~\cite{HRIS_Mag_all,Tsinghua_RIS_Tutorial, Swindlehurst_CE}), and since all channels fluctuate due to fast fading conditions, the system's long-term performance is captured through the ergodic averages of the mutual information over the fading distribution. For convenience, we mathematically define the ergodic sum-MI performance for the considered MIMO-MAC-RIS system as:
\begin{align}
    \label{eq:I(Q,Phi)}
    &I\left(\{\bQ_m\}_{m\in{\cal S}},\{\bPhi_k\}_{k=1}^K\right) 
    \\ \nonumber
    &\triangleq \ex\left[\log\det\left(    
    \vec I_\nr + \sum_{m\in {\cal S}}\bG_{{\rm tot},m} \bQ_m \bG_{{\rm tot},m}^\dagger
    \right)\right],
\end{align}
where the averaging is over the channel matrices $\bG_{{\rm tot},m}$'s for ${\cal S} \subseteq \{1,2,\ldots,M\}$. In this expression, we explicitly show the dependences on the input covariance matrices $\bQ_m$'s, as well as the phase configurations matrices $\bPhi_k$'s for all RISs. The above expression represents the ergodic sum-MI for the set ${\cal S}$ of TX users being active, while all others being silent.

\subsection{MIMO-MAC without RISs}
\Revadd{In the absence of RISs (hence, setting $\bPhi_k=\bZero_{\ns}$ $\forall$$k$ in~\eqref{eq:I(Q,Phi)})}, the capacity region of a pure MIMO-MAC system has been well analyzed in the past \cite{Cheng1993_GaussianMAC_ISI_CapacityRegion, Tse1998_GaussianMAC1_PolymatroidStructure, Vishwanath2001_OptimumMAC, Goldsmith2003_CapacityLimitsMIMO, Yu2004_IterativeMIMOMAC}
and consists of all rate multiplets $\{R_1,R_2,\ldots,R_M\}$ such that it holds:
\begin{align}
\label{eq:cap_region_def}
    \sum\limits_{m\in {\cal S}} 
    R_m \leq I\left(\{\bQ_m\}_{m\in{\cal S}}\right)
\end{align}
for all sets ${\cal S}\subseteq \{1,2,\ldots,M\}$; in this expression, for simplicity, we do not include sum-MI's dependence on $\bPhi_k$'s, since they are all equal to the identity matrix. The above capacity region is achieved with Gaussian inputs at each TX, where each covariance matrix $\bQ_m$ at the each $m$-th TX is constrained such that $\Tr{\bQ_m}\leq \nt$.

The covariance matrices that correspond to the points on the boundary of the capacity region can be obtained by maximizing the following functional with respect to all $\bQ_m$'s~\cite{Yu2004_IterativeMIMOMAC}:
\begin{align}\label{eq:functional_pure}
    &{\cal L}\left(\{\bQ_m\}_{m=1}^M,\boldmu\right)
    \\ \nonumber
    &\triangleq
    \mu_M I\left(\{\bQ_m\}_{m=1}^M\right) 
+\sum_{\ell=1}^{M-1}\left(\mu_{\ell}-\mu_{\ell+1}\right) 
    I\left(\{\bQ_m\}_{m=1}^\ell\right) 
\end{align}
for all possible non-negative vectors $\boldmu\triangleq[\mu_1,\mu_2,\ldots,\mu_M]$ of relative priorities given to each of the TXs and for a fixed sequential interference cancellation order, where it was assumed without loss of generality that $\mu_1\geq \mu_2\geq \ldots \mu_M$. The optimization is performed given the statistical knowledge of the channel matrices (covariances matrices) appearing in \eqref{eq:all_correlations}. 

\subsection{MIMO-MAC Assisted by RISs}
The above analysis can be readily generalized in the presence of RISs. In this case, to attain the boundary points of the capacity region, the ergodic sum-MI needs to be also maximized over all $\bPhi_k$ matrices. Hence, we have the following proposition.
\begin{Proposition}[{\bf MIMO-MAC-RIS Ergodic Capacity Region}]
\label{prop:cap_region} 
Let the channel matrices $\bG_{{\rm tot},m}$'s for $m=1,2,\ldots,\nue$ be composed as in \eqref{eq:Gtot}, including the diagonal matrices $\bPhi_k$'s for $k=1,2,\ldots,K$ with the tunable reflection coefficients for each of the $K$ RISs, and the input signal covariance matrices $\bQ_m$'s for each of the $M$ TX antenna arrays. The MIMO-MAC-RIS capacity region consists of all rate multiplets $\{R_1,R_2,\ldots,R_M\}$ such that it holds using \eqref{eq:I(Q,Phi)}:
\begin{align}
\label{eq:cap_region_prop}
    \sum\limits_{m\in {\cal S}} 
    R_m \leq I\left(\{\bQ_m\}_{m\in{\cal S}},\{\bPhi_k\}_{k=1}^K\right)
\end{align}
for all sets ${\cal S} \subseteq \{1,2,\ldots M\}$ with $\Tr{\bQ_m}\leq \nt$ $\forall$$m$. The matrices $\bQ_m$'s and $\bPhi_k$'s corresponding to the borders of this region are obtained by maximizing the following functional: 
\begin{align}\label{eq:Lagrangian_mu_cap}
    &L_M\!\left(\{\bQ_m\}_{m=1}^M,\{\bPhi_k\}_{k=1}^K,\boldmu\right)\triangleq  
    \mu_M I\left(\{\bQ_m\}_{m=1}^M,\{\bPhi_k\}_{k=1}^K\right)\nonumber
\\ 
&+\sum_{\ell=1}^{M-1}\left(\mu_{\ell}-\mu_{\ell+1}\right) 
    I\left(\{\bQ_m\}_{m=1}^\ell,\{\bPhi_k\}_{k=1}^K\right)
\end{align}
for all possible non-negative vectors $\boldmu$ as in \eqref{eq:functional_pure} of relative priorities given to each TX and for a fixed sequential interference cancellation order, where we have again assumed without loss of generality that $\mu_1\geq \mu_2\geq \ldots \mu_M$ and $\sum_{\ell=1}^M\mu_\ell=1$. 
\end{Proposition}
\begin{proof}
The proof is a trivial extension of the MIMO-MAC results in \cite{Cheng1993_GaussianMAC_ISI_CapacityRegion, Tse1998_GaussianMAC1_PolymatroidStructure, Vishwanath2001_OptimumMAC, Goldsmith2003_CapacityLimitsMIMO, Yu2004_IterativeMIMOMAC} in the absence of RISs.
\end{proof}
To make progress along the lines of the latter proposition, the quantity $I(\{\bQ_m\}_{m=1}^M,\{\bPhi_k\}_{k=1}^K)$ needs to be evaluated for fixed input covariance matrices $\bQ_m$'s and RIS reflection matrices $\bPhi_k$'s. To do this, we will generalize the recent results from \cite{Moustakas2023_RIS} to obtain novel asymptotic closed-form expressions for the MIMO-MAC-RIS ergodic MI in the limit of large number of RIS elements $\ns$, while at the same time the numbers $\nt$ of TXs' antennas and the number $\nr$ of the RX antennas also grow at the same rate. Then, in the next section, we will focus on optimizing $\bPhi_k$'s in order to maximize ${\cal L}(\{\bQ_m\}_{m=1}^M,\{\bPhi_k\}_{k=1}^K,\boldmu)$ as a function of $\boldmu$, regarding the ergodic average of $I(\{\bQ_m\}_{m=1}^M,\{\bPhi_k\}_{k=1}^K)$ in \eqref{eq:I(Q,Phi)}. For concreteness, we will assume that the channels and covariances from all TXs participate in the optimization, a result which can easily be generalized to all subsets ${\cal S}$ of TXs.
\begin{Proposition}[{\bf Asymptotic Mean and Variance of $I(\{\bQ_m\}_{m=1}^M,\{\bPhi_k\}_{k=1}^K)$}]
\label{prop:ergMI} 
Let the channel matrices $\bG_{{\rm tot},m}$'s take the form of \eqref{eq:Gtot} including the RIS reflection diagonal matrices $\bPhi_k$'s, the matrices $\bG_{r,k}$'s for the channels between RX and each $k$-th RIS, and, for each $m$-th TX, the matrices $\bG_{dm}$ and $\bG_{t,km}$'s for the direct channel to RX and the channel matrices from the $k$-th RIS, respectively. Assume also that the above channel matrices have zero-mean complex Gaussian elements with covariances given by \eqref{eq:Grk_cov}, \eqref{eq:Gd_cov}, and \eqref{eq:Gtk_cov}, respectively. When $\nt, \nr, \ns\to\infty$ with fixed ratios $\beta_r\triangleq\nr/\nt$ and $\beta_s\triangleq\ns/\nt$, then the quantity $I(\{\bQ_m\}_{m=1}^M,\{\bPhi_k\}_{k=1}^K)$ normalized per TX antenna element can be expressed as follows:
\begin{align}\label{eq:S0}
& C_M\left(\{\bQ_m\}_{m=1}^M,\{\bPhi_k\}_{k=1}^K\right) \triangleq \frac{\ex[I\left(\{\bQ_m\}_{m=1}^M,\{\bPhi_k\}_{k=1}^K\right)]}{\nt}
\nonumber  \\ %
&=\frac{1}{\nt}\log\det \left(\bI_\nr  + \tilde{\bR} \right)
\nonumber  \\ %
    &+\frac{1}{\nt} \sum_{k=1}^K\sum_{m=1}^M \log\det \left(\bI_{\ns} +  
t_{1k}r_{2km}\bSigma_{km}\right)
\nonumber \\%
  &+\frac{1}{\nt}\sum_{m=1}^\nue\log\det \left(\bI_\nt + 
  \bQ_m \tilde{\bT}_m  \right)
   \nonumber \\
 &- \sum_{m=1}^M\left(r_{dm} t_{dm}+\sum_{k=1}^K\left(r_{1km}t_{1k}+r_{2km}t_{2km}\right)\right).
\end{align}
In the above, the matrices $\tilde{\bR}$, $\tilde{\bT}_m$, and $\bSigma_{km}$ are defined as: 
\begin{align}
    \tilde{\bR}&\triangleq 
    \sum_{m=1}^\nue\left( r_{dm} \bR_{dm} + \sum_{k=1}^\nris r_{1km}\bR_{k}\right),
    \label{eq:R_tilde}\\
    \tilde{\bT}_m&\triangleq 
    t_{dm}\bT_{dm}+  \sum_{k=1}^\nris t_{2km}\bT_{km},
    \label{eq:T_tilde}\\
    \bSigma_{km} &\triangleq 
    \bS_{t,km}^{1/2}\bPhi_k^\dagger\bS_{r,k}\bPhi_k\bS_{t,km}^{1/2}
    \label{eq:Sigma_k_initial},
\end{align}
where the quantities $t_{dm}$, $r_{dm}$, $t_{1k}$, $r_{1km}$, $t_{2km}$,  and $r_{2km}$ can be evaluated by solving the fixed point equations as below:
\begin{align}
\nonumber
    t_{dm} &=\frac{1}{\nt}\Tr{\left(\bI_\nr + \tilde{\bR} \right)^{-1}\bR_{dm}},
  \\ \nonumber
    t_{1k} &=\frac{1}{\nt}\Tr{\left(\bI_\nr   + \tilde{\bR}  \right)^{-1}\bR_{k}},\\
    \nonumber
    r_{dm} &= \frac{1}{\nt} \Tr{\left(\bI_\nt + \bQ_m
    \tilde{\bT}_m
    \right)^{-1}\bQ_m\bT_{dm}}, 
  \\ \nonumber
    r_{2km} &= \frac{1}{\nt} \Tr{\left(\bI_\nt + 
    \bQ_m \tilde{\bT}_m
    \right)^{-1}\bQ_m\bT_{km}},
\end{align}    
\begin{align} 
   \nonumber
  r_{1km}& = \frac{r_{2km}}{\nt} \Tr{\left(\bI_{\ns} +  
t_{1k}r_{2km}\bSigma_{km}\right)^{-1}  \bSigma_{km}},
  \\
  t_{2km} &= \frac{t_{1k}}{\nt} \Tr{\left(\bI_{\ns} +  
t_{1k}r_{2km}\bSigma_{km}  \right)^{-1} \bSigma_{km}}.
\label{eq:fp_eqs}
\end{align}
In addition, the variance of $I(\{\bQ_m\}_{m=1}^M,\{\bPhi_k\}_{k=1}^K)$ takes the limiting form:
\begin{align}\label{eq:Var(I)}
 {\rm Var}(I(\{\bQ_m\}_{m=1}^M,\{\bPhi_k\}_{k=1}^K))\triangleq-\log\det({\bf \Lambda}),
\end{align}
where the $(2M+4MK)$-dimensional matrix $\bLambda$ appears in the appendix in \eqref{eq:Vmat_def}. 
\end{Proposition}
\begin{proof}
The proof appears in the Appendix.
\end{proof}
\begin{remark}[Capacity Achieving Covariance Matrices]
Using the analysis in~\cite{Moustakas2007_MIMO1}, it can be shown that the asymptotic ergodic sum MI $C(\{\bQ_m\}_{m=1}^M,\{\bPhi_k\}_{k=1}^K)$ is maximized over each covariance matrix $\bQ_m$ when it is diagonalized by the same unitary matrix as ${\tilde \bT}_m$. This is a result of the fact that, each $\bQ_m$ only appears in \eqref{eq:S0} as a product $\bQ_m{\tilde \bT}_m$, and that, since the values of the parameters $\{r\}$ and $\{t\}$ are solutions of the saddle-point equations \eqref{eq:fp_eqs}, their first-order variation vanishes. \Revadd{Hence, for each $m=1,2,\ldots, M$ in the basis of ${\tilde \bT}_m$, $\bQ_m$ takes the form $\bQ_m=diag\{\bq_m\}$, where $\bq_m$ a vector of elements such that $q_{am}\geq 0$ $\forall$$a=1,2,\ldots,\nt$. For each $m$, the vector $\bq_m$ is chosen to maximize the corresponding term in the third line of \eqref{eq:S0}, namely  $\log\det(\bI_\nt +   \bQ_m \tilde{\bT}_m)$, subject to the total power constraint $\Tr{\bQ_m}= \nt \rho_m$. Using \cite{Fiedler1971_BoundsDetSumHermitianMatrices}, it can be shown that the solution to this maximization problem is given by the statistical waterfilling~\cite{Moustakas2007_MIMO1}, so that:
\begin{align}
    \sup_{\Tr{\bQ_m}=\nt\rho_m}\left[\log\det \left(\bI_\nt +   \bQ_m \tilde{\bT}_m \right)\right]=\sum_{a=1}^{\nt} \left[\log\left(\frac{\tau_{am}}{\lambda_m}\right)\right]_+
\end{align}
where $[x]_+=0$ if $x\leq 0$ and $[x]_+=x$ if $x>0$. In this expression, $\boldtau_m\triangleq[\tau_{1m}\,\tau_{a2m}\,\ldots\,\tau_{\nt m}]$ is the vector including the eigenvalues of 
$\tilde{\bT}_m$ and $\lambda_m$ is a parameter obtained from the following waterfilling constraint:
\begin{equation}
    \frac{1}{\nt}\sum_{a=1}^\nt \left[\frac{1}{\lambda_m} -\frac{1}{\tau_{am}}\right]_+ = \rho_m.
\end{equation}
}

\end{remark}
\begin{remark}[Decoupling Effects]\label{rem:Decoupling}
In the above expression, we note that in the asymptotic limit a number of important simplifications take place. First, we see that the matrices $\bPhi_k$'s, appearing in the first line of \eqref{eq:S0} through the quantities $\bSigma_{km}$, depend only on the incoming and outgoing covariance matrices $\bS_{r,k}$'s and $\bS_{t,km}$'s, and not on other covariance matrices of the problem. Furthermore, as also noted in \cite{Moustakas2023_RIS}, the $\bPhi_k$'s of the different RISs are decoupled, since they appear in separate $\log\det(\cdot)$ terms. This allows for separate optimization of the phases of each matrix $\bPhi_k$. \Revadd{The underlying reason for this feature is the fact that, as seen in \eqref{eq:basic_channel_eq}, the RIS phase configuration matrices appear ``sandwitched'' between independent channel matrices. Since the phases of these matrices are independent, it is easy to see that they can be optimized independently. In the asymptotic limit, the effect of all other phase configuration matrices enters into the optimization of a given phase matrix $\bPhi_k$ through the scalar quantities $r_{1km}$, $t_{1k}$, $r_{2km}$, $t_{2km}$, $r_{dm}$, and $t_{dm}$ in~\eqref{eq:fp_eqs}, which only depends on all other phase matrices and in an aggregate manner.} In addition, the $\bSigma_{km}$'s each corresponding to a different $m$-th TX, also appear in separate $\log\det(\cdot)$ terms. As we shall see below, this will enable faster optimization of the $\bPhi_k$ matrices.
\end{remark}
\begin{remark}[Asymptotic MIMO-MAC-RIS Ergodic Capacity Region]
Combining Propositions~\ref{prop:cap_region} and~\ref{prop:ergMI}, we obtain the ergodic capacity region per antenna in the asymptotic limit of large numbers of TXs/RX antennas and RIS reflecting elements.
\end{remark}
\begin{remark}
The variance of $I(\{\bQ_m\}_{m=1}^M,\{\bPhi_k\}_{k=1}^K)$ provides a metric for its variability. For example, we see that. while this metric is of order $O(\nt)$, its variance is of order unity. Furthermore, since all higher moments can be shown to vanish in the large $\nt$ limit, the distribution of $I(\{\bQ_m\}_{m=1}^M,\{\bPhi_k\}_{k=1}^K)$ can be shown to be asymptotically Gaussian. Although this is beyond the scope of this work, using the theoretical framework developed in  \cite{Tse1998_GaussianMAC1_PolymatroidStructure}, one can obtain the outage capacity region for the above problem. This outage capacity region can be calculated more easily in the asymptotic limit, by noting that the joint probability distribution of $I(\{\bQ_m\}_{m=1}^M,\{\bPhi_k\}_{k=1}^K)$, for different sets ${\cal S}$ can be shown, using tools from statistical physics \cite{Moustakas2003_MIMO1}, to be jointly Gaussian. 
\end{remark}

\section{MIMO-MAC-RIS Capacity Optimization}\label{sec:MI_Optimization}
After analyzing the asymptotic ergodic sum-MI performance of MIMO-MAC-RIS systems in the previous section, we will now discuss the optimization of the ergodic sum-MI with respect to the phase configuration of the multiple RISs, namely the elements of all $\bPhi_k$'s. The purpose of the optimization is twofold. First, it results to the maximum ergodic sum-capacity of such systems, which is a metric indicating the maximum total throughput for the case of $\nue$ TXs and $\nris$ RISs. In addition, it provides the boundaries of the ergodic MIMO-MAC capacity region in the presence of multiple RISs, as shown in Proposition~\ref{prop:cap_region}. As mentioned in the previous section, we will focus only in the case where there is no need to optimize the signal covariance matrices $\bQ_m$'s, for example, when the covariance matrices at all $M$ TXs, i.e., $\bT_{km}$'s and $\bT_{dm}$'s, are identity matrices. Hence, following the system model in~\eqref{eq:Gtot} and the geometry depicted in Fig.~\ref{fig:system_model}, and using the capacity region in~\eqref{eq:Lagrangian_mu_cap} as well as the closed-form asymptotic sum-MI expression in~\eqref{eq:S0}, our Optimization Problem (OP) formulation is mathematically expressed as follows:
\begin{align*} 
\begin{split}
    \mathcal{OP}_1: \,\, \max_{\{\bPhi_k\}_{k=1}^\nris} &
    \,\,\Bigg[\mu_M C_M\left(\{\bPhi_k\}_{k=1}^K\right)+
    \\ 
    &\hspace{0.2 cm}\left.\sum_{\ell=1}^{M-1}\left(\mu_{\ell}-\mu_{\ell+1}\right) 
    C_\ell\left(\{\bPhi_k\}_{k=1}^K\right)\right]
    \\ 
    & \hspace{-0.4cm}\text{s.t.} \quad \lvert [\bPhi_k]_{n,n} \rvert = 1  \, \, \forall k,n.
\end{split}
\end{align*}
As mentioned in Remark \ref{rem:Decoupling}, the ergodic sum-MI in~\eqref{eq:S0}, using~\eqref{eq:R_tilde}--\eqref{eq:fp_eqs}, is simpler, when compared to the corresponding  exact expression~\eqref{eq:I(Q,Phi)}, since the 
RISs' phase configuration matrices $\bPhi_k$'s appear in separate logarithms. This property simplifies the solution of $\mathcal{OP}_1$; interestingly, the asymptotic sum-MI optimization can be performed separately for each $\bPhi_k$, i.e., for each $k$-th RIS.

To simplify the exposition of the methodology, we will next present only the case of the sum capacity, namely, when $\mu_M=1$ and $\mu_1=\ldots=\mu_{M-1}=0$. In this case, to solve the resulting $\mathcal{OP}_1$ form, we deploy an alternating optimization approach~\cite{J:alternating_minimization}, according to which, at each algorithmic iteration, we initially fix the values of the variables $r_{1km}$, $t_{1k}$, $r_{2km}$, $t_{2km}$, $r_{dm}$, and $t_{dm}$ in~\eqref{eq:fp_eqs} and maximize the sum capacity over the $\bPhi_k$'s. As previously mentioned, it is possible to separate the process into  independent ones over different $\bPhi_k$'s. Hence, the multi-RIS phase configuration design problem can be made more simple as follows. For each $k$-th RIS, making use of~\eqref{eq:S0}, the following OP needs to be solved:
\begin{align*} 
\begin{split}
    \mathcal{OP}_2: \max_{\bPhi_k}\sum_{m=1}^M &\log\det \left(\bI_{\ns} + t_{1k}r_{2km}\bPhi_k^\dagger\bS_{r,k}\bPhi_k\bS_{t,km} \right)
\\    \hspace{0.4cm}
\text{s.t.} \quad &\lvert [\bPhi_k]_{n,n} \rvert = 1  \, \, \forall k,n.
\end{split}
\end{align*}
Once $\mathcal{OP}_2$ is solved for each of the $\nris$ RISs, we substitute all obtained phase matrices into~\eqref{eq:fp_eqs} to calculate the parameters $r_{1km}$, $t_{1k}$, $r_{2km}$, $t_{2km}$, $r_{dm}$, and $t_{dm}$. We may repeat these two steps iteratively, until the all parameters converge to their optimal values, or until the number of iterations exceeds its maximum value.

We will now present two related ways to obtain the optimum $\bPhi_k$'s, each solving one of the $\nris$ separate $\mathcal{OP}_2$ problems, with both being based on the gradient ascent approach. The first, in Section~\ref{sec:Analytical_Solution} that follows, is a semi-optimal algorithm, which, while not converging at a maximum, is less computationally intensive and performs well at low signal angular spreads, as will be demonstrated numerically in the following Section~\ref{sec:Numerical_Results}. Subsequently, in Section \ref{sec:Numerical_Solution}, we will describe the full gradient ascent approach solving $\mathcal{OP}_2$.

\subsection{Semi-Optimal Optimization over $\bPhi_k$'s}
\label{sec:Analytical_Solution}  
To better understand the form of the sum-capacity-optimal solution for $\bPhi_k$'s, we will first focus on a MIMO-MAC-RIS system with one RIS, dropping for convenience the RIS numbering index $k$. In addition, we will analyze the limit of vanishing angle spread, for which the matrices $\bS_{t,m}$'s and $\bS_{r}$ are effectively of unit rank, corresponding to a single incoming plane wave to the RIS from each $m$-th TX and a single outgoing plane wave from the RIS, respectively. In~\cite{Moustakas2023_RIS}, we showed that the RIS phase configuration solutions for this vanishing angle spread approximation case can very accurately agree with the full optimization analysis for general angle spreads. This will be similarly confirmed for this paper's MIMO-RIS-MAC focus in the next numerical results section. Hence, let us commence by expressing $\bS_{t,m}=\ns \bu_m\bu_m^\dagger$ $\forall$$m$ and $\bS_{r}=\ns \bv\bv^\dagger$,
where the unit-norm vectors $\bu_m$ and $\bv$ correspond to the wave vectors of the incoming signal from the $m$-th TX and the outgoing signal to the RX and have components, each having the elements:
\begin{align}
    \label{eq:S_eigenvector}
    \left[\bu_m\right]_n = \frac{1}{\sqrt{\ns}} e^{i\bq_{t,m}\bx_n}, \quad
    \left[\bv\right]_n = \frac{1}{\sqrt{\ns}} e^{i\bq_{r}\bx_n},
\end{align}
where $\bx_n$ represents the location of the $n$-th RIS element, and $\bq_{t,m}$ and $\bq_{r}$ are the incoming wave-vectors from the $m$-th Tx to the RIS and the outgoing wave vector from the RIS to the RX, respectively. For this case where the angle spread is vanishing, the matrices $\bSigma_m$'s can be expressed as follows:
\begin{align}
\label{eq:unit_rank_Sigma_m}
    \bSigma_m= N_s^2 |\kappa_{m}(\bPhi)|^2 \bu_{m}\bu^\dagger_{m},
\end{align}
where the scalar parameter $\kappa_{m}$ is given by:
\begin{align}
\label{eq:kappa_m}
\kappa_{m}(\bPhi)&\triangleq\bv^\dagger\bPhi\bu_{m}=\frac{1}{\ns}\sum_{n=1}^{\ns} e^{i\left(\phi_n-\Delta\bq_m\bx_n\right)}.
\end{align}
with $\Delta\bq_m\triangleq\bq_{r}-\bq_{tm}$. By substituting~\eqref{eq:S_eigenvector} into~\eqref{eq:unit_rank_Sigma_m} and then the resulting expression into the previously derived asymptotic ergodic sum-MI formula, the second term in~\eqref{eq:S0} can be written as follows (for the case of a single RIS):
\begin{align}\label{eq:C_M}
  D_M(\bPhi) \triangleq  \sum_{m=1}^M \log\left(1 + N_s^2t_{1m}r_{2m}|\kappa_m(\bPhi)|^2\right).
\end{align}
\begin{remark}
In the case of only one TX in the system, i.e., when $M=1$, the summation in~\eqref{eq:C_M} diminishes to one term including $\kappa_1(\bPhi)$, and it can be optimized over this parameter when the modulus of $\kappa_1(\bPhi)$ is maximal, i.e., when $\phi_n=\Delta\bq\bx_n$ $\forall$$n$; this has been recently shown in~\cite{Moustakas2023_RIS}. However, in the general MIMO-MAC-RIS case with $M>1$ TXs, each of the phase components $\phi_n$ needs to be chosen to optimize all $\kappa_m(\bPhi)$'s simultaneously, as shown from the $M$ terms included in~\eqref{eq:C_M}. It is clear that, the more separated the quantities $\Delta\bq_m$'s for different values of $m$ are, the higher the competition of the terms in \eqref{eq:C_M}'s summation will be, and hence, the smaller the values of $D_M(\bPhi)$ in~\eqref{eq:S0} will be. 
\end{remark}

To optimize over the RIS phase configuration $\bPhi$ (i.e., solve $\mathcal{OP}_2$ for the vanishing angle spread case), we perform gradient ascent on the above functional. It can be shown that, at each $i$-th algorithmic iteration, the reflection coefficient at each $n$-th RIS element in the diagonal of $\bPhi^{(i)}$ is updated as follows: 
\begin{align}\label{eq:GD_iteration_simple}
    &\phi_n^{(i)}=\phi_{n}^{(i-1)}
    \\ \nonumber
    &+\!\epsilon\sum_{m=1}^M{\rm Im}\left(\! \frac{\ns t_{1m}r_{2m}\kappa_m\left(\bPhi^{(i-1)}\right) e^{i\left(\Delta \bq_m \bx_n-\phi_n^{(i-1)}\right)} }{1\!+\!N_s^2t_{1m}r_{2m}\left|\kappa_m\left(\bPhi^{(i-1)}\right)\right|^2}   \!\right)\!\!,
\end{align}
where $\epsilon$ is a parameter chosen to make the algorithm converge fast. When the above iteration converges, e.g., at the $\mathcal{I}$-th iteration, we use the obtained matrix $\bPhi^{(\mathcal{I})}$ to compute the fixed point values of the parameters in~\eqref{eq:fp_eqs}, which are then fed back into~\eqref{eq:C_M} until the whole algorithm converges. 

Note that the same procedure can be employed to obtain the borders of the capacity region using~\eqref{eq:Lagrangian_mu_cap} and combining Propositions~\ref{prop:cap_region} and~\ref{prop:ergMI}. In particular, in this case, for fixed $\boldmu$, we adopt gradient ascent to maximize the following functional with respect to the RIS phase configuration matrix $\bPhi$: 
\begin{align}\label{eq:cap_mu12}
  D_M(\bPhi) \triangleq  \sum_{\ell=1}^M (\mu_\ell-\mu_{\ell+1})\!\sum_{m=1}^\ell \log\left(1 + N_s^2t_{1m}^\ell r_{2m}^\ell |\kappa_m(\bPhi)|^2\right)\!,
\end{align}
where, for simplicity, we have set $\mu_{M+1}=0$ in~\eqref{eq:Lagrangian_mu_cap}. In the above, the superscripts on the parameters $t_{1m}^\ell$ and $r_{2m}^\ell$ are included to specify that these correspond for each $\ell$-value to the fixed parameters of the equations~\eqref{eq:fp_eqs} for the case of $\ell$ TXs.

\subsection{Optimal Solution for $\bPhi_k$'s}\label{sec:Numerical_Solution}   
In the general case of arbitrary values for the angular spread, we need to evaluate the full gradient of the functional objective appearing in $\mathcal{OP}_2$ with respect to each of the RIS phase configuration matrices. Similar to the treatment in Section~\ref{sec:Analytical_Solution}, we next focus on a single $\bPhi$. Clearly, solving $\mathcal{OP}_2$ is computationally more complex, due to the necessity to invert a large $\ns\times\ns$ matrix. Specifically, at each $i$-th algorithmic iteration, the reflection coefficient at each $n$-th RIS element in the diagonal of $\bPhi^{(i)}$ needs to be updated as follows: 
\begin{align}\label{eq:GD_iteration_full}
    &\phi_n^{(i)}=\phi_{n}^{(i-1)}
    \\ \nonumber 
    &+\!\epsilon\!\sum_{m=1}^M{\rm Im}
    \left[\!
    \left(\bI_{\ns}\! + \!t_{1m}r_{2m}\left(\bPhi^{(i-1)}\right)^\dagger\bS_{r}\bPhi^{(i-1)}\bS_{t,m} \right)^{\!-1}\!
    \right]_{n,n}\!\!,
\end{align}
where the values of the parameters $t_{1m}$ and $r_{2m}$ are fed from the previous iteration. In principle, and as mentioned above, we should perform the above iteration until it converges, for example, at the $\mathcal{I}$-th iteration, and then use the resulting $\bPhi^{(\mathcal{I})}$ matrix (for every RIS) to re-evaluate the parameters $t_{1m}$ and $r_{2m}$ using ~\eqref{eq:fp_eqs}. However, since the equations in~\eqref{eq:fp_eqs} also involve the evaluation of the inverse of the matrix appearing above, it is more computationally efficient to jointly perform the optimization of the parameters and the RIS phase configuration matrix. The procedure for obtaining the optimal solution of the sum-MI metric in $\mathcal{OP}_1$, i.e., when $\mu_M=1$ and $\mu_1=\ldots=\mu_{M-1}=0$, for multiple RISs is summarized in Algorithm~\ref{alg:loc}. We have found this approach to be numerically stable, especially when compared to the algorithm of~\cite{Zhang_Capacity}, and similar ones, that perform algorithmic iterations per RIS element. \Revadd{ Also, we have tested and found the same convergence values with several initial conditions (for example, all zeros or random initial conditions).} The overall algorithmic steps for solving the same problem via the semi-optimal approach of Section~\ref{sec:Analytical_Solution} can be obtained in a similar manner.
\begin{algorithm}[!t]
    \caption{MIMO-MAC-RIS Sum-MI Optimizing $\bPhi_k$'s}
    \label{alg:loc}
    \begin{algorithmic}[1]
        \renewcommand{\algorithmicrequire}{\textbf{Input:}}
        \renewcommand{\algorithmicensure}{\textbf{Output:}}
        \REQUIRE $\epsilon$, $\delta$, $\mathcal{I}$, $\bR_{dm}$, $\bT_{dm}$, $\bR_{k}$, $\bS_{r,{k}}$, $\bS_{t,km}$, and $\bT_{km}$ $\forall$$k,m$. 
        \STATE Initialize $r_{1km}^{(0)}=t_{1k}^{(0)}=r_{2km}^{(0)}=t_{2km}^{(0)}=r_{dm}^{(0)}=t_{dm}^{(0)}=0$ and $\bPhi_k^{(0)}=\bI_\ns$ $\forall$$k,m$.
        \FOR{$i=1,2,\ldots,\mathcal{I}$ }
        \FOR{$k=1,2,\ldots,\nris$ }
                \STATE Set $\mathbf{A}_{km}^{(i)}\!=\!t_{1km}^{(i-1)}r_{2km}^{(i-1)}\bS_{t,km}\left(\!\bPhi_k^{(i-1)}\!\right)^\dagger\bS_{r,k}\bPhi_k^{(i-1)}$ $\forall$$m$. 
                \STATE Compute $\mathbf{B}_{km}^{(i)}\!=\!\left(\bI_{\ns}+\mathbf{A}_{km}^{(i)}\right)^{-1}$
                .\STATE Use matrix $\mathbf{B}_{km}^{(i)}$ to update $r_{1km}^{(i)}$ and $t_{2km}^{(i)}$ using~\eqref{eq:fp_eqs}, as well as $t_{dm}^{(i)}$, $t_{1km}^{(i)}$, $r_{dm}^{(i)}$, and $r_{2km}^{(i)}$.
                \FOR{$n=1,2,\ldots,\ns$ }
                \STATE Set $\phi_{k,n}^{(i)}=\phi_{k,n}^{(i-1)}+\!\epsilon\!\sum_{m=1}^M{\rm Im}\left[\mathbf{B}_{km}^{(i)}\right]_{n,n}$.
                \ENDFOR
        \ENDFOR
                \IF{$\left\lVert\bPhi_{k}^{(i)}-\bPhi_{k}^{(i-1)}\right\rVert+  
                \left|r_{1km}^{(i)}-r_{1km}^{(i-1)}\right|+\left|t_{2km}^{(i)}-t_{2km}^{(i-1)}\right|+\left|t_{dm}^{(i)}-t_{dm}^{(i)}\right|+\left|t_{1km}^{(i-1)}-t_{1km}^{(i-1)}\right|+\left|r_{dm}^{(i)}-r_{dm}^{(i-1)}\right|+\left|r_{2km}^{(i)}-r_{2km}^{(i-1)}\right|
                <\delta$ }
                    \STATE Output the optimized $\bPhi_k^{(i)}$ and break.
        \ENDIF
        \ENDFOR
        \STATE Output each optimized $\bPhi_k^{(\mathcal{I})}$.
    \end{algorithmic}
\end{algorithm}

\section{Numerical Results and Discussion}\label{sec:Numerical_Results}
\begin{figure}[!t]
\centering
\includegraphics[width=1.00\columnwidth]{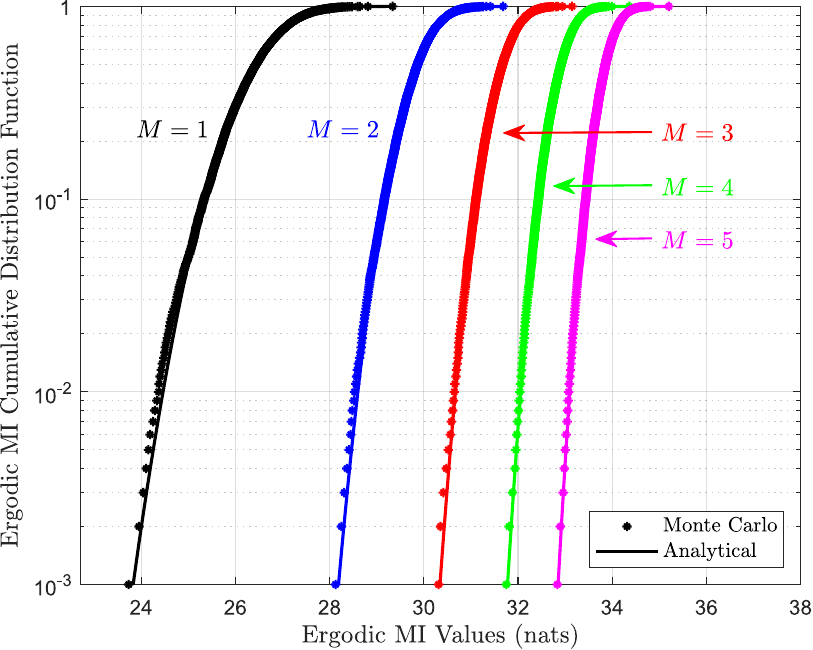}
\caption{The cumulative distribution distribution of the sum MI in nats with $\nris=2$ RISs and with $\nue=1$ (black), $2$ (blue), $3$ (red), $4$ (green), and $5$ (magenta) TXs, respectively. We have used $\nt=8$ and $\nr=4$ antennas and $\ns=400$ elements per RIS, and $\rho=10$ dB. The solid curves correspond to a Gaussian approximation using the mean and variance obtained analytically in Proposition~\ref{prop:ergMI}, and are compared with curves generated via Monte Carlo simulations. For simplicity, we have assumed all channels to be uncorrelated and have set $\bPhi_k=\bI_{400}$ $\forall$$k$. It can be observed that, for increasing $\nue$, the gain in the median throughput as well as the variance (related to the average slope of the curves) is progressively decreasing. It is also evident that the agreement between the theoretical and numerically generated distributions is particularly good down to $10^{-3}$ outage probability.}
\label{fig:MI_CDFs}
\end{figure}

In this section, we present numerically evaluated results on the statistics of the sum-MI performance for the considered MIMO-MAC-RIS system, between multiple TX arrays and a single RX array, using both the ``Semi-Optimal'' and the ``Optimal'' optimization approaches presented in the previous Sections~\ref{sec:Analytical_Solution} and~\ref{sec:Numerical_Solution}, respectively. We also showcase the accuracy of our analytic results for the mean and variance of the sum MI by comparing them with equivalent Monte-Carlo-generated channel matrix instantiations. To focus on the effects of the presence of RISs, in all performance evaluation figures that follow we have neglected the direct paths $\bG_{dm}$'s between the TXs and RX, hence, we have set $\gamma_{dm}=0$ $\forall$$m=1,2,\ldots,M$. In addition, we have assumed that all TX antenna arrays as well as the RX antenna array exhibit uncorrelated fading, and set a constant SNR for all TXs. In all but Fig.~\ref{fig:MI_CDFs} that both follow, we have assumed a specific geometry of the relative location of the multiple RISs relative to the RX antenna array and the TX arrays, and have kept a number of parameters fixed. In addition, we have fixed the incoming and outgoing mean elevation angles of the signal waves to and from the RISs, respectively, so that geometric optical reflections not to be possible, and thus, the optimization of the reflectivity of the RISs to be necessary. Our detailed simulation parameters are summarized in Table~\ref{table1}. 
\begin{figure}[!t]
	\centering
	\includegraphics[width=\columnwidth]{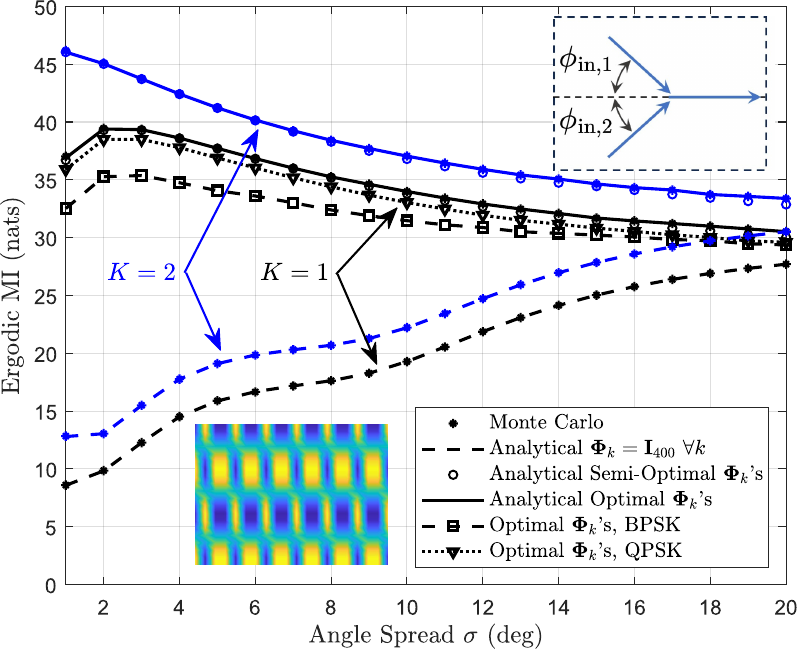}
	  \caption{The ergodic sum-MI performance in nats per channel use for $M=2$ TXs and for $K=1$ and $2$ RISs versus the angle spread $\sigma$ in degrees. All parameter values of this simulation setup appear in Table~\ref{table1}, except for the incoming signal azimuth angles from each TX to the RISs, which are set as $\phi_{{\rm in},1}=45^{\circ}$ and $\phi_{{\rm in},2}=-45^{\circ}$, as can be seen in the upper inset. In the case of two RISs, we have kept for simplicity the same parameter values, essentially locating the second RIS at the mirrored location of the first with respect to the plane of the TXs and the RX. The ``o'' points correspond to the $\bPhi_k$'s obtained using the algorithm of Section~\ref{sec:Analytical_Solution}, while the solid curves correspond to the optimized $\bPhi_k$'s presented in Section~\ref{sec:Numerical_Solution}. We see that they practically coincide with each other and with the Monte Carlo simulations. It is also depicted that the benefit of having two RISs (blue curves) serving both TXs is relatively minor, at least as compared with the large performance increase of the presence of one RIS compared to none (dashed curves). \Revadd{We have also added two curves depicting the ergodic sum MI for the case of a single RIS when its phases are $1$-bit quantized to BPSK values (i.e., $e^{i \phi}=\pm 1$) and $2$-bit quantized QPSK values (i.e., $e^{i \phi}=(\pm 1\pm i)/\sqrt{2}$). It is shown that, in both cases, the optimization gain is significant, with the QPSK case being practically identical to the fully optimal case. Similar behavior had been obtained when two RISs are present, but we have not included the curves so as not to clutter the figure.} Finally, the lower inset figure depicts the optimal phase distribution on the RIS \Revadd{for the continuous case} when the angle spread is equal to $\sigma=5^{\circ}$.}
		\label{fig:MI_AS_2UE}
\end{figure}
\begin{table}[!t]
\caption{Simulation Parameters used in Figs.~\ref{fig:MI_AS_2UE}--\ref{fig:cap_region}.}
\label{table1}
\centering
\begin{tabular}{|c|c||c|c|}
\hline
\textbf{Parameter} &\textbf{Value} & \textbf{Parameter} &\textbf{Value}\\
\hline \hline
Carrier frequency $f_c$ & $2.5$GHz & Wavelength $\lambda$& $12$ cm \\
\cline{1-2} \cline{3-4}
Direct SNR $\gamma_{dm}$ & $0$ dB & SNR $\rho$ & $10$ dB \\
\cline{1-2} \cline{3-4}
TX antennas $\nt$ & $8$ & RX antennas $\nr$ & $4$  \\
\cline{1-2} \cline{3-4}
RIS elements $\ns$ & $400$ & RIS inter-element distance & $6$ cm  \\
\hline
Elevation $\theta_{\rm in}$ & $30^{\circ}$ &Elevation $\theta_{out}$ & $70^{\circ}$ \\
\hline
\end{tabular}
\vspace{0.1cm}

\end{table}

\begin{figure}[!t]
	\centering
	\includegraphics[width=\columnwidth]{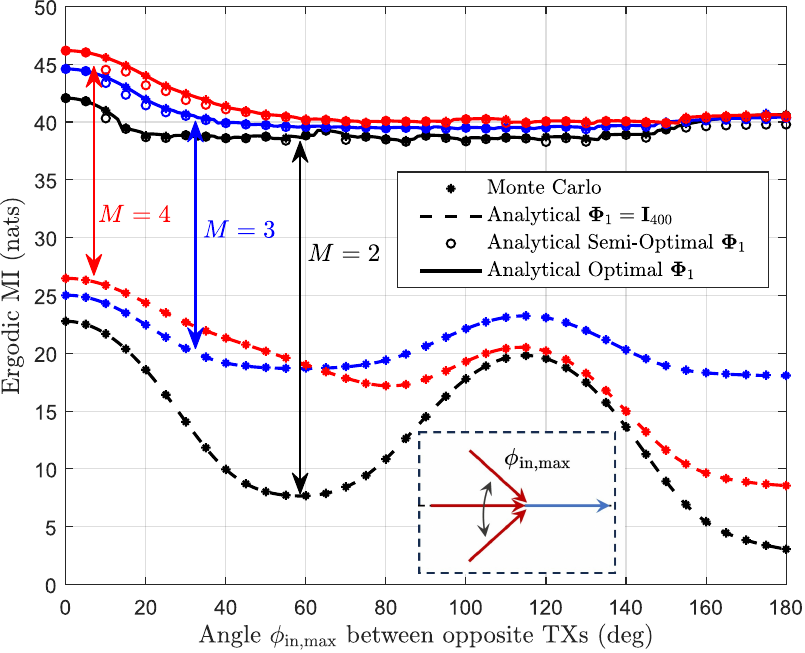} 
 \caption{The ergodic sum-capacity performance in nats per channel use in the presence of $K=1$ RIS for $\nue=2$ (black), $3$ (blue), and $4$ (red) TXs. All incoming and outgoing signals were considered to have angle spread $\sigma=4^{\circ}$. The remaining simulation parameters take values from Table~\ref{table1}. For the cases where $\nue>1$, the TXs have incoming azimuth angles that are equidistant with maximum angle equal to the one plotted on the $x$-axis of the plot. For concreteness, we have included an inset which shows the azimuth angles for the case of $\nue=3$. The solid lines correspond to the optimized $\bPhi_1$, while the dashed ones correspond to the semi-optimal approach described in Section~\ref{sec:Analytical_Solution}. The dotted lines correspond to Monte Carlo simulations, while the lower curves depict the sum-MI without any optimization, i.e., for $\bPhi_1=\bI_{400}$. It is evident that, for increasing $\nue$, the relative gain is diminishing, Also, for increasing azimuth distance between TXs, it is demonstrated that the capacity gains are decreasing. This happens because the optimum $\bPhi_1$ has difficulties satisfying all TXs effectively. }
\label{fig:MI_phi_1234UE}
\end{figure}

\begin{figure}[!t]
\centering
\includegraphics[width=\columnwidth]{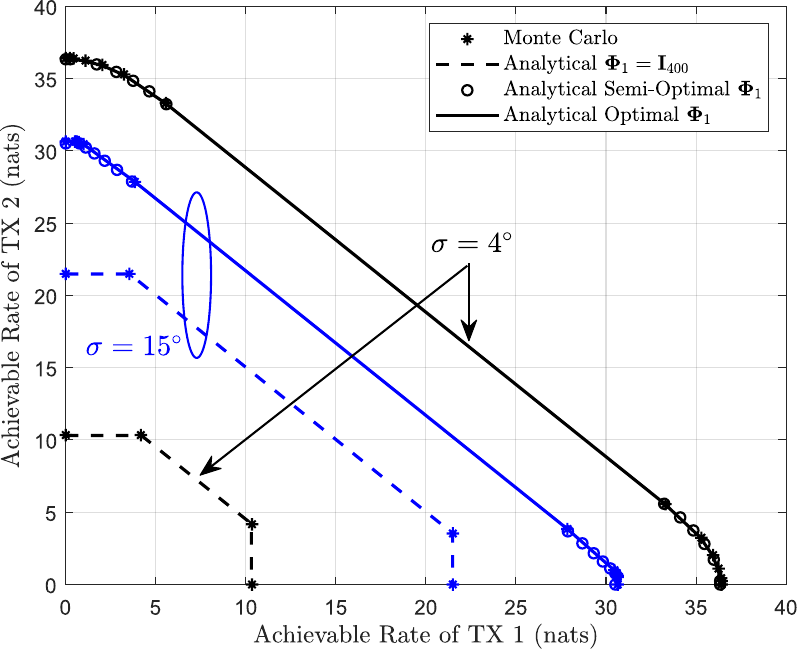}
\caption{Ergodic capacity region for the case of $\nue=2$ TXs and $K=1$ RIS, with incoming signal azimuth angles from each TX set to $\phi_{{\rm in},1}=45^{\circ}$ and $\phi_{{\rm in},2}=-45^{\circ}$, considering the angle spread values $\sigma=4^{\circ}$ (black) and $\sigma=15^{\circ}$ (blue). All other parameter values are included in Table~\ref{table1}. The upper two curves correspond to the capacity region boundary obtained by optimizing $\bPhi_1$ for $0\leq\mu_1\leq 1$ and $\mu_2=1-\mu_1$ in~\eqref{eq:cap_mu12}. It is clear that (especially for $\sigma=4^{\circ}$) the boundary is not a pentagon, in contrast to the lower two curves for which $\bPhi_1=\bI_{400}$. The fact that the lower angle spread has a wider capacity region can be expected from Fig.~\ref{fig:MI_AS_2UE}. The solid curves correspond to the optimal capacity region boundaries calculated using the full optimization approach of Section~\ref{sec:Numerical_Solution}. Finally, the rate pairs evaluated from Monte Carlo simulations using the $\bPhi_1$ obtained by numerically optimizing expression~\eqref{eq:Lagrangian_mu_cap} for $\mu_1=0:0.1:1$ and $\mu_2=1-\mu_1$ are illustrated.}
\label{fig:cap_region}
\end{figure}

The ergodic sum-MI performance in nats per channel use (npcu) as a function of the angle spread $\sigma$ in degrees, for $\nue=2$ TX arrays and both $K=1$ and $2$ RISs, is illustrated in Fig.~\ref{fig:MI_AS_2UE}. The mean azimuth angle of the impinging signal to the single RX is fixed to $\phi_{\rm out}=0^{\circ}$, while the corresponding incoming mean azimuth angles from the two TXs take two values, namely $\phi_{{\rm in},1}= 45^{\circ}$ and $\phi_{{\rm in},2}=-45^{\circ}$, as illustrated in the upper inset figure. In the baseline case of $\bPhi_k=\bI_{400}$ $\forall$$k$ (dashed curves), where no optimization of the RIS phase matrices is performed, the throughput increases with angle spread, which is a direct consequence of the increasing diversity. In contrast, the solid curves, corresponding to the ergodic sum-capacity for optimized $\bPhi_k$'s, are maximal at low angle spreads. This phenomenon is due to the significant beamforming gain resulting from the optimized coherent reflections from the RISs along the optimal directions determined from the correlation matrices of the two TXs. 
To illustrate the optimal solution for the RIS phase configuration matrices, the lower inset figure depicts the cosine of the optimal phase at each element of an RIS. 
It is interesting to note the relatively slow variation of the optimal phase-angle on the RIS. This suggests that the information necessary for the optimization of the RIS phase matrix may not scale with the number of its elements.  
\Revadd{It can be also observed from the figure that near optimal behavior can be obtained even when the phases of the RIS elements are quantized. We depict the ergodic sum-MI performance when the diagonal elements of phase matrix (for the case of one RIS for simplicity) are quantized with $1$ bit (taking the Binary phase-shift keying (BPSK) values $e^{i\phi}=\pm 1$) and $2$ bits (with Quadrature Phase Shift Keying (QPSK) values $e^{i\phi}=(\pm 1\pm i)/\sqrt{2}$. In fact, in the case of the $2$-bit quantization, the performance is nearly identical to the phase-continuous optimal case.}
Furthermore, the blue curves depict the sum MI in the presence of two RISs. In order not to introduce additional parameters, we assume, for simplicity, that the corresponding angle spreads are equal and the incoming and outgoing mean angles are mirror images of each other. We conclude that the relative gain of adding the first RIS compared to that of adding a second one is significantly higher. 
Finally, it also demonstrated in the figure that the less computationally heavy, semi-optimal optimization method of the phase matrices (``o'' points) presented in Section~\ref{sec:Analytical_Solution} produces practically identical performance as the full optimization approach (solid curves) described in Section~\ref{sec:Numerical_Solution}, while all analytic results agree very well with equivalent ones obtained by means of Monte Carlo simulations.
\Revadd{Concluding, the small gain one gets from the second RIS is essentially an SNR gain. In contrast, one could envision to optimize each RIS for reflection of the signal from one of the TXs, oblivious to the other. This would presumably provide roughly the same sum MI, but without having to deal with multiple pilots, the resulting control overhead~\cite{RIS_CC_v1}, and the corresponding pilot contamination.}

In Fig.~\ref{fig:MI_phi_1234UE}, we investigate how effective the optimization of RISs can be for increasing numbers of TXs, and particularly when the incoming signals' angles vary significantly. For simplicity, we have assumed a MIMO-MAC-RIS system with single RIS and $\nue$ TX antenna arrays communicating with a single RX array. For the case with $\nue>1$, the mean incoming azimuth angle of each TX was assumed to differ by $\phi_{{\rm in},\max}/(\nue-1)$. As an example, the inset on the bottom of the figure showcases how the mean incoming azimuth angles are distributed in the angular space when $\nue=3$. For each value of $\nue=2,3,$ and $4$, we plot in Fig.~\ref{fig:MI_phi_1234UE} the sum-MI as a function of $\phi_{{\rm in},\max}$, assuming that the angle spread for each of the incoming and outgoing beams is $\sigma=4^{\circ}$ around the mean direction, with all other values for the simulation parameters set as indicated in Table~\ref{table1}. It can be seen that, when the phase matrix $\bPhi_1$ is optimized, the sum-capacity increases with increasing $\nue$. However, the relative gain decreases with $\nue$, and beyond $\nue=3$, the relative gains of the sum-capacity with $\nue$ become negligible, at least for significant spreading between the incoming signals, $\phi_{{\rm in},\max}$. For completeness, we also include the non-optimized sum-MI values, which are all significantly lower that their optimized equivalents. As in the previous figure, the results from Monte Carlo simulations are almost indiscernible from the analytic results, both optimized and with a unit phase configuration matrix. It is also depicted that the semi-analytic curves (``o'' points), obtained using the approach presented in Section~\ref{sec:Analytical_Solution}, are quite close to the fully optimized ones via the approach in Section~\ref{sec:Numerical_Solution} (solid curves), except at very small and large $\phi_{{\rm in},\max}$ values. \Revadd{It is finally noted that this fact provides insight in the positioning of the RISs so as to serve TXs with at most small angular separation. In addition, it posits that, beyond that angular distance, one needs to use multiple RISs, a result that also contributes towards the RIS dimensioning challenge.}

The previous figure suggests that adding additional TXs to the system decreases the relative gain of each additional TX. Hence, one needs to assess the achievable rate pairs when optimizing the phase configuration matrix for each RIS. Therefore, in Fig.~\ref{fig:cap_region}, we evaluate the capacity region for a $2$-user MIMO-MAC-RIS system, where we have considered, for simplicity, a single RIS. This numerical evaluation is performed for two values of the incoming signals' angle spreads, namely, for low $\sigma=4^{\circ}$ (blue) and moderate $\sigma=15^{\circ}$ (green), with the two mean azimuth angles of the two TXs being $\phi_{{\rm in},1}=45^{\circ}$ and $\phi_{{\rm in},2}=-45^{\circ}$, as appearing in the inset of Fig.~\ref{fig:MI_AS_2UE}. All other simulation parameters take values as appearing in Table~\ref{table1}. The border of the capacity regions, when taking into account the optimization of $\bPhi_1$, appears in solid lines, resulting from the optimization of~\eqref{eq:cap_mu12}. We observe that, in contrast to the usual pentagon capacity regions for the non-optimized $\bPhi_1$ appearing with dashed lines, the capacity region including optimization over $\bPhi_1$ has a continuous slope. This happens due to the fact that, every pair $(\mu_1,\mu_2)$ with priority values for the considered two TXs corresponds to a different optimal $\bPhi_1$. Once again, it is shown that the agreement between Monte Carlo simulations and our analytic results is remarkable, both for the proposed semi-optimal and fully optimal approaches.

Finally, to showcase the accuracy of the Gaussian approximation for the outage capacity presented in Proposition~\ref{prop:ergMI}, we compare in Fig.~\ref{fig:MI_CDFs} the distribution of the sum-MI for $\nris=2$ RISs and various numbers $\nue$ for the TXs generated using Monte Carlo simulations, with the Gaussian approximation of the distribution using the analytical results for the mean in~\eqref{eq:S0} and the variance in~\eqref{eq:Var(I)} of the sum-MI. For simplicity, we have set $\bPhi_k=\bI_{400}$ for both $k=1$ and $2$, and assumed uncorrelated channels. The numerically generated and theoretical distributions agree extremely well with each other .

\section{Conclusions and Future Work}
\label{sec:conclusion}
In this paper, we have obtained the asymptotic mean and variance of the sum-MI for MIMO-MAC systems when multiple transmitters as well as multiple RISs are present, for the relevant case of correlated channels. To evaluate these quantities we have employed tools from statistical physics, such as the method of replicas. Although these results are valid for large RIS element and antenna numbers, we have found that even for modest-sized RISs and arrays, the Gaussian approximation obtained using only these two cumulants is extremely accurate even at very low outages. The above conclusions are especially relevant when the MIMO-MAC-RIS system experiences fast-fading conditions, so that the phase-configurations of the multiple RISs can only be optimized in a timely manner using statistical channel knowledge. In addition, we showed that, when we optimize the RISs' phase matrices for fixed priorities of the multiple TXs, we can obtain the capacity region for the MIMO-MAC-RIS problem. 

Using the ergodic sum-MI performance as a metric of the total system performance, it was shown that, optimizing over the phase profiles of the RISs leads to significant throughput benefits, which however are diminished for growing angular separation between the transmitting users, as well as increasing anglespread of the ingoing and outgoing signals at the RISs. Furthermore, we observed that, as the number of TXs increases beyond a certain point, the relative gains from optimizing the multiple RISs become small. As a result, additional RISs are necessary to serve increasing numbers of TXs. Moving forward, it is important to analyze the potential effects of electromagnetic interactions among the elements of the multiple RISs in such smart wireless environments, as well as other realistic effects and aspects of RISs, such as the polarization and their control \Revadd{(determined also by their hardware architecture and resulting operation modes~\cite{zhi2022active,HRIS_Mag_all})}, which may significantly affect the success of the RIS technology for 6G wireless networks. \Revadd{Furthermore, interesting inroads may be taken using this paper's methodology towards the analysis of the multi-user multi-RIS broadcast channel using duality \cite{chen2023fundamental}.}

\appendix
\label{app:proof_ergMI}
In this appendix, we extend the approach developed in~\cite{Moustakas2023_RIS} to obtain the analytical expressions for the asymptotic mean and variance of the normalized sum-MI $C(\{\bQ_m\}_{m=1}^M,\{\bPhi_k\}_{k=1}^K)$ presented in Proposition~\ref{prop:ergMI}. To this end, we first define the scalar quantity:
\begin{equation}
    {\cal Z}\triangleq\det\left(\bI_\nr+\sum_{m=1}^M\bG_{{\rm tot},m}\bQ_m\bG_{{\rm tot},m}^\dagger\right)^{-1},
\end{equation}
and its Moment Generating Function (MGF) $g(\nu)\triangleq\ex\left[{\cal Z}^\nu\right]$.
Then, the normalized ergodic MI appearing in the first row of~\eqref{eq:S0} can be obtained from the first derivative of $g(\nu)$ as follows:
\begin{equation}\label{eq:c=loggnu}
C\left(\{\bQ_m\}_{m=1}^M,\{\bPhi_k\}_{k=1}^K\right) =- \nt^{-1}\left.\frac{d\log g(\nu)}{d\nu}\right|_{\nu=0^+}.
\end{equation}
In addition, the variance of the MI metric is deduced as:
\begin{equation}\label{eq:VarI=loggnu''}
    \text{Var}\left(I(\{\bQ_m\}_{m=1}^M,\{\bPhi_k\}_{k=1}^K)\right)=\left.\frac{d^2\log g(\nu)}{d\nu^2}\right|_{\nu=0^+}.
\end{equation}
To be able to derive analytic formulas for the above metrics, the limit for $\nt\to\infty$ of their definitions need to be computed.  

Following~\cite[Assumption~1]{Moustakas2023_RIS}, which have been shown to hold in a related setting to this paper \cite{Guerra2002_ThermodynamicLimitSG, Talagrand2006_ParisiFormula,Moustakas2003_MIMO1,Tanaka2002_ReplicasInCDMAMUD}, we will use the expressions of $g(\nu)$ evaluated below for replica index values $\nu\in\ZZ^+$, and analytically continue them to values of $\nu$ close to $\nu=0^+$. By using~\cite{Moustakas2007_MIMO1}, it is straightforward to derive:
\begin{align}\label{eq:Z_power}
    {\cal Z}^\nu = &\ex\left[
    e^{\frac{1}{2}\sum_{m=1}^\nue\Tr{\bY_m^\dagger\bG_{dm}^\dagger  \bZ-\bZ^\dagger \bG_{dm}\bQ_m\bY_m }}
\right. 
\\ \nonumber
&\!
    \times e^{\frac{1}{2}\!\!\sum_{k=1}^\nris\sum_{m=1}^\nue\!\!\Tr{\bV_{km}^\dagger\bG_{t,km}\bY_m-\bY_m^\dagger \bQ_m\bG_{t,km}^\dagger\bW_{km}}}
\\ \nonumber
&\left.\!
         \times e^{-\frac{1}{2}\!\!\sum_{k=1}^\nris\sum_{m=1}^\nue\!\!\Tr{\bZ^\dagger\bG_{r,k}\bPhi_k\bV_{km}+\bW_{km}^\dagger\bPhi^\dagger_k\bG_{r,k}^\dagger\bZ}}
        \right]\!\!,
\end{align}
The expectation symbol above denotes integration over the complex matrices 
$\bZ$, $\bY_m$, $\bV_{km}$, and $\bW_{km}$ ($k=1,2,\ldots,\nris$, $m=1,\ldots,\nue$), 
which are assumed to be complex Gaussian, each with zero mean and variance equal to $2$, resulting to the following MGF expression:
\begin{align}
\label{eq:gnu1}
&g(\nu) =  \ex\left[ e^{ -\frac{1}{4\nt}\sum_{m=1}^\nue \Tr{\bY_m^\dagger\bQ_m\bT_{d,m}\bY_m\bZ^\dagger\bR_{dm}\bZ} } 
\right.
\\ \nonumber
&\!\times 
e^{\frac{1}{4\nt} \sum_{k=1}^\nris\sum_{m=1}^\nue\Tr{\bZ^\dagger\bR_{k}\bZ\bW_{km}^\dagger\bPhi_k^\dagger\bS_{r,k}\bPhi_{k}\bV_{km} } }\!\!.
\\ \nonumber
&\left.\!\times 
e^{-\frac{1}{4\nt} \sum_{k=1}^\nris\sum_{m=1}^\nue \Tr{\bY_m^\dagger\bQ_m\bT_{km}\bY_m\bV_{km}^\dagger\bS_{t,km}\bW_{km}}} \right]\!\!.
\end{align}
The integration over the Gaussian channel matrices has resulted to an exponent with quartic terms in the Gaussian random variables. By employing the so-called Hubbard-Stratonovich transformation~\cite[Identity~1]{Moustakas2023_RIS}, we can integrate over the complex matrices $\bZ$, $\bY_m$, $\bV_{km}$, and $\bW_{km}$, by decomposing the above quartic terms into quadratic ones through the introduction of the $\nu\times\nu$ matrices
${\mathbfcal T}_{dm}$, ${\mathbfcal R}_{dm}$,
${\mathbfcal T}_{1km}$, ${\mathbfcal R}_{1km}$ and
${\mathbfcal T}_{2km}$, ${\mathbfcal R}_{2km}$, with $k=1,2,\ldots,\nris$ and $m=1,2,\ldots,\nue$.
As a result, \eqref{eq:gnu1} can be re-written as follows:
\begin{align}
\label{eq:gnu2} %
g(\nu)=\int e^{-{\cal E }}d\mu(\{\bTcal, \bRcal\}),
\end{align}
where $d\mu(\{\bTcal, \bRcal\})$ represents the integration measure over the above-introduced $\nu\times\nu$ matrices and ${\cal E}$ can be expressed as:
\begin{align}
 &{\cal E} \triangleq \log\det\!\left(\!\bI_{\nr\nu}\!+\!\sum_{m=1}^\nue\!\left(\bR_{dm} \!\outer\! \bRcal_{dm}\!+\!
  \sum_{k=1}^\nris \bR_{k} \outer \bRcal_{1km}\right)\!\!\right)
\nonumber \\
&+\!\sum_{m=1}^\nue\log\det \left(\!\bI_{\nt\nu}\!+\!\bQ_m\!\left(\bT_{dm} \!\outer\! \bTcal_{dm}\!+\!
  \sum_{k=1}^\nris \bT_{km}\!\outer\!\bTcal_{2km}\right)\!\!\right)
\nonumber \\
 &+\!\sum_{m=1}^M\sum_{k=1}^K \log\det\left(\!\bI_{\nu\ns}\!+\! 
\bTcal_{1km}\bRcal_{2km}\!\outer\! \bPhi_k^\dagger\bS_{r,k}\bPhi_k\bS_{t,km}  \right) \nonumber
\\&-\!\nt\sum_{m=1}^M\Tr{\bTcal_{dm} \bRcal_{dm}\!+\!\sum_{k=1}^\nris \left(\bTcal_{1km} \bRcal_{1km} + \bTcal_{2km} \bRcal_{2km}\right) }
\label{eq:S_full}
\end{align}
with the symbol $\outer$ indicating the direct product of matrices.

Subsequently, the integral in \eqref{eq:gnu2} is calculated in the asymptotic limit of large $\nt$, $\nr$, and $\ns$. To this end, we assume that the analytic continuation of $g(\nu)$ to the real value $\nu=0^+$ in both \eqref{eq:c=loggnu} and \eqref{eq:VarI=loggnu''} can be interchanged with the limit $\nt\to\infty$~\cite[Assumption 2]{Moustakas2023_RIS}, and then by deforming the integral-contours of all matrix-elements of $\{\mathbfcal{T}, \mathbfcal{ R}\}$ to traverse a saddle-point of ${\cal E}$~\cite{Moustakas2003_MIMO1, Bender_Orszag_book}. Since the $\nu$ replicas are \textit{a priori} identical with each other, following ~\cite[Assumption 3]{Moustakas2023_RIS}, the matrices $\{\mathbfcal{T}, \mathbfcal{ R}\}$ appearing in \eqref{eq:S_full} should be expected to have rotational symmetry in the space of continuous replica rotations and therefore they are proportional to $\bI_\nu$. As a result, we have:
\begin{align}
    \left.{\mathbfcal T}_{dm}\right|_{{\rm saddle}\mbox{-}{\rm point}}\!\!=\!t_{dm}\bI_\nu &\mbox{,} \left.{\mathbfcal R}_{dm}\right|_{{\rm saddle}\mbox{-}{\rm point}}\!\!=\!r_{dm}\bI_\nu, \nonumber
    \\ 
    \left.{\mathbfcal T}_{akm}\right|_{{\rm saddle}\mbox{-}{\rm point}}\!\!=\!t_{akm}\bI_\nu &\mbox{,} \left.{\mathbfcal R}_{akm}\right|_{{\rm saddle}\mbox{-}{\rm point}}\!\!=\!r_{akm}\bI_\nu,
\end{align}
The parameters $r_{dm}$, $t_{dm}$, $r_{akm}$, and $t_{akm}$, are evaluated through the saddle-point equations, which produce the fixed-point equations in \eqref{eq:fp_eqs} with unique solutions \cite{Taricco2008_MIMOCorrelatedCapacity}. 
Denoting as ${\cal E}_0$ the saddle-point value of ${\cal E}$, the function $g(\nu)$ in \eqref{eq:gnu2} can be re-expressed as follows:
\begin{equation}\label{eq:g_nu}
    g(\nu)=e^{-{\cal E}_0} \int e^{-({\cal E}-{\cal E}_0)} d\mu(\{{\mathbfcal{T}, \mathbfcal{R}}\}.
\end{equation}
It can be seen that the saddle-point ${\cal E}_0$ takes the value $\nu \nt C$, where $C$ appears in \eqref{eq:S0}. 

To evaluate the MI variance from the previous expression, we perform a Taylor-expansion  of ${\cal E}$ around its value at the saddle-point, integrating exactly the quadratic term, while neglecting higher-order terms as subleading perturbations, which are $O(\nt^{-1})$. This yields the following expression for ${\cal E}$ that includes ${\cal E}_2$ which will be defined in the sequel:
\begin{eqnarray}\label{eq:S_expansion}
{\cal E}={\cal E}_0+\nt\left({\cal E}_2+ \sum_{\ell=3}^\infty {\cal E}_\ell\right),
\end{eqnarray}
\begin{align}
    &{\cal E}_2 = \nonumber 
     -\!\sum_{m=1}^M\sum_{k=1}^K\Tr{\delta{\mathbfcal T}_{1km} \delta{\mathbfcal R}_{1km}\!+\!\delta{\mathbfcal T}_{2km} \delta{\mathbfcal R}_{2km}}
    \\ \nonumber 
    &-\sum_{m=1}^M\Tr{\delta{\mathbfcal T}_{dm} \delta{\mathbfcal R}_{dm}}\\ \nonumber 
    &+
    \sum_{m=1}^M\sum_{k=1}^K[{\bf M}_{12}]_{k,m}^{km} \Tr{\delta{\mathbfcal T}_{1km} \delta{\mathbfcal R}_{2km}}      
    \\ \nonumber   
    &+\frac{1}{2}\! 
    \sum_{m,m'=1}^M\sum_{k,k'=1}^K [{\bf M}_{1r}]_{k,m}^{k'm'}
    \Tr{ \delta\mathbfcal{R}_{1km}\delta\mathbfcal{R}_{1k'm'}} 
    \\ \nonumber   
    &+\frac{1}{2}\! 
    \sum_{m=1}^M\sum_{k,k'=1}^K [{\bf M}_{2t}]_{k,m}^{k'm} 
    \Tr{ \delta\mathbfcal{T}_{2km}\delta\mathbfcal{T}_{2k'm}}
    \\ \nonumber
    &+\frac{1}{2}\!
    \sum_{m=1}^M\sum_{k=1}^K \left([{\bf M}_{2r}]_{k,m}^{km}
    \Tr{ \delta\mathbfcal{R}_{2km}\delta\mathbfcal{R}_{2km}}\right.\\ \nonumber &\left.\hspace{1.9cm}+ [{\bf M}_{1t}]_{k,m}^{km} 
    \Tr{ \delta\mathbfcal{T}_{1km}\delta\mathbfcal{T}_{1km}} \right)
    \\ \nonumber
    &+\sum_{m,m'=1}^M\sum_{k=1}^K [{\bf M}_{d1r}]_{k,m}^{m'}
    \Tr{ \delta\mathbfcal{R}_{1km}\delta\mathbfcal{R}_{dm'}} 
    \\ \nonumber
    &+\sum_{m=1}^M\sum_{k=1}^K [{\bf M}_{d2t}]_{k,m}^m
    \Tr{ \delta\mathbfcal{T}_{2km}\delta\mathbfcal{T}_{dm}}
    \\ \nonumber
    &+\frac{1}{2}\!\sum_{m=1}^M [{\bf M}_{dt}]_{m,m}
    \Tr{ \delta\mathbfcal{T}_{dm}^2} \\ 
    &+\frac{1}{2}\!\sum_{m,m'=1}^M [{\bf M}_{dr}]_{m,m'}
    \Tr{ \delta\mathbfcal{R}_{dm}\delta\mathbfcal{R}_{dm'}},
\end{align}
where $\delta {\mathbfcal R}_{1km} = {\mathbfcal R}_{1km}-\left.{\mathbfcal R}_{1km}\right|_{{\rm saddle}\mbox{-}{\rm point}}$, etc., are $\nu\times\nu$ matrices. The matrices ${\bf M}_{1r}$, etc., above are defined as follows:
\begin{align}
    &[{\bf M}_{1r}]_{k,m}^{k'm'} \triangleq -\frac{1}{\nt} \Tr{ \overline{{\bf R}} {}^{-1}  {\bf R}_{km}\overline{{\bf R}} {}^{-1} {\bf R}_{k'm'}}, 
    \nonumber \\ \nonumber
    &[{\bf M}_{2r}]_{k,m}^{k'm'} \triangleq -\delta_{kk'}\delta_{mm'}\frac{t_{1k}^2}{\nt} \Tr{ \left(\overline{{\bf S}}_{km}\right)^{-2}  {\bf \Sigma}_{km}^2 },
    \\ \nonumber
    &[{\bf M}_{1t}]_{k,m}^{k'm'} \triangleq -\delta_{kk'}\delta_{mm'}\frac{r_{2km}^2}{\nt} \Tr{ \left(\overline{{\bf S}}_{km}\right)^{-2}  {\bf \Sigma}_{km}^2 },
    \\ \nonumber
    &[{\bf M}_{2t}]_{k,m}^{k'm'} \triangleq  -\frac{\delta_{mm'}}{\nt}
    \\ \nonumber
    &\quad\quad\times\Tr{ \left(\overline{{\bf T}}_m\right)^{-1}  {\bf Q}_m{\bf T}_{mk}\left(\overline{{\bf T}}_m\right)^{-1} {\bf Q}_m{\bf T}_{mk'}},\\
        \nonumber
    &[{\bf M}_{12}]_{k,m}^{k'm'} \triangleq \delta_{kk'}\delta_{mm'}\frac{1}{\nt} \Tr{ \left(\overline{{\bf S}}_{km}\right)^{-2}  {\bf \Sigma}_{km} },\\
    \nonumber
    &[{\bf M}_{1dr}]_{k,m}^{m'} \triangleq -\frac{1}{\nt} \Tr{ \overline{{\bf R}} {}^{-1}  {\bf R}_{km}\overline{{\bf R}} {}^{-1} {\bf R}_{dm'}},\\
    \nonumber
    &[{\bf M}_{2dt}]_{k,m}^{m'}\triangleq -\frac{\delta_{mm'}}{\nt}
    \\ \nonumber
    & \quad\quad\times\Tr{ \left(\overline{{\bf T}}_m\right)^{-1}  {\bf Q}_m{\bf T}_{km}\left(\overline{{\bf T}}_m\right)^{-1} {\bf Q}_m{\bf T}_{dm}},
    \\ \nonumber
    &[{\bf M}_{dt}]_{m,m'} \triangleq -\frac{1}{\nt} \Tr{ \left(\left(\overline{{\bf T}}_m\right)^{-1}  {\bf Q}_m{\bf T}_{dm}\right)^2},\\
    &[{\bf M}_{dr}]_{m,m'}\triangleq  -\frac{1}{\nt} \Tr{ \overline{{\bf R}} {}^{-1}  {\bf R}_{dm} \overline{{\bf R}} {}^{-1}  {\bf R}_{dm'}},
    \end{align}
%
where we have used the following notation:
\begin{align}
    \overline{{\bf R}} &\triangleq 
    {\bf I}_{\nr}+\tilde{\bf R},\quad\overline{{\bf T}}_m \triangleq 
    {\bf I}_{\nt}+{\bf Q}_m\tilde{\bf T}_m, \nonumber
    \\ 
    \overline{{\bf S}}_{km} &\triangleq 
    {\bf I}_{\ns}+t_{1k}r_{2km}\bSigma_{km}
\end{align}

After integrating over the matrices $\delta\mathbfcal{T}$ and $\delta\mathbfcal{R}$ following the methodology of \cite{Moustakas2007_MIMO1}, the moment-generating function of the sum-mutual information takes the following form:
\begin{equation}
    g(\nu)=e^{-{\cal E}_0} 
    \det\left({\bf \Lambda}\right)^{-\frac{\nu^2}{2}},
\end{equation}
with the $(4M\nris+2M)$-dimensional matrix ${\bf \Lambda}$ is given by:
\begin{align}\label{eq:Vmat_def}
    {\bf \Lambda} = \left[\begin{array}{cccccc} 
    {\bf M}_{dt} & 0 & {\bf M}_{2dt}^T & 
    -\bI_{\nue} & 0 & 0
    \\   0 & {\bf M}_{1t} & 0 & 
    0 & -\bI_{\nue\nris} & {\bf M}_{12}
    \\
    {\bf M}_{2dt} & 0 & {\bf M}_{2t} & 
    0 & 0 & -\bI_{\nue\nris}
    \\
    -\bI_\nue & 0 & 0 & 
    {\bf M}_{dr} & {\bf M}_{1dr}^T & 0
    \\
    0 & -\bI_{\nue\nris} & 0 & 
    {\bf M}_{1dr} & {\bf M}_{1r} & 0
    \\
    0 & {\bf M}_{12} & -\bI_{\nue\nris} & 
    0 & 0 & {\bf M}_{2r}
    \end{array}\right].
\end{align}
After some manipulations, we obtain \eqref{eq:Var(I)}. It can be shown that, following the methodology in \cite{Moustakas2003_MIMO1, Moustakas2007_MIMO1}, all higher cumulants of the sum-MI are negligible when $\nt\to \infty$, thus resulting to an asymptotically Gaussian distribution for this metric.

\bibliographystyle{IEEEtran}
\bibliography{IEEEfull,wireless,references}

\end{document}